\documentclass[reprint,amssymb, amsmath, aip,cha,twocolumn]{revtex4-1}
\usepackage{epsfig,subfigure,float,wrapfig,float,color}
\usepackage{graphicx, setspace}
\usepackage[colorlinks=true,linkcolor=blue]{hyperref}%
\expandafter\ifx\csname package@font\endcsname\relax\else
 \expandafter\expandafter
 \expandafter\usepackage
 \expandafter\expandafter
 \expandafter{\csname package@font\endcsname}%
\fi
\begin{document}
\title{ Dynamics of compressional Mach cones in a strongly coupled complex plasma}
\author{P. Bandyopadhyay$^{1,}$\renewcommand{\thefootnote}{\alph{footnote}} \footnote{Electronic mail: pintu@ipr.res.in},  R. Dey$^1$, Sangeeta Kadyan$^2$ and Abhijit Sen}
\affiliation{
$^1$ Institute for Plasma Research, Bhat, Gandhinagar-382428, India\\
$^2$ Department of Physics, Maharshi Dayanand University, Rohtak- 124001, India.}

%\ead{pintu@mpe.mpg.de}
%\date{\today}
%#####################################################################################
\begin{abstract}
Using a Generalised-Hydrodynamic (GH) fluid model we study the influence of strong coupling induced modification of the fluid compressibility on the dynamics of compressional Mach cones in a dusty plasma medium. A significant structural change of lateral wakes for a given Mach number and Epstein drag force is found in the strongly coupled regime. With the increase of fluid compressibility, the peak amplitude of the normalised perturbed dust density first increases and then decreases monotonically after reaching its maximum value. It is also noticed that the opening angle of the cone structure decreases with the increase of the compressibility of the medium and the arm of the Mach cone breaks up into small structures in the velocity vector profile when the coupling between the dust particles increases.     
\end{abstract}
%\pacs{52.27.Lw, 52.35.Mw, 47.40.-x}
%\submitto{\NJP}
\maketitle
\section{Introduction} 
A complex plasma (or dusty plasma), consisting of electrons, ions and micron-sized dust grains, can have distinctly different behaviour from the usual two component electron-ion plasmas due to the presence of highly charged dust particles.  When micron or sub-micron sized dust particles are introduced in a conventional plasma  they get  charged by  collecting electrons and ions from the background. The charged dust component can become strongly coupled if its potential energy exceeds its kinetic energy. The strength of interaction is characterized by the Coulomb coupling parameter $\Gamma(= Q_d^2/4\pi\epsilon_0dT_d \text{exp}(-d/\lambda_D))$, where $Q_d$, $d$, $T_d$ and $\lambda_D$ are the dust charge, inter-particle distance, dust temperature and plasma Debye length, respectively. When $\Gamma$ exceeds a critical value $\Gamma_c$ the dust component can crystallize and behaves like an ordered solid \cite{Ikezi_1986, Thomas_1994}. For a given set of plasma parameters, if the temperature of the dust particles is increased then the crystal can melt and vaporise like in an ordinary fluid. Hence, a complex plasma provides an excellent medium to investigate phase transitions occuring as one moves from a strongly coupled regime to a weakly coupled regime. In that context a topic of considerable interest is how such phase transitions influence the existence and propagation characteristics of linear and nonlinear modes as well as vortex and wake structures. Wake structures arise behind an object moving in a fluid and can become a cone shaped shock structure (a Mach cone) if the particle velocity exceeds the sound speed in the medium.  Mach cones are very familiar in the field of gas dynamics \cite{Bond_1965} and solid matter \cite{Cheng_1994}. They have also been extensively studied theoretically \cite{Havnes1,  Havnes2, Dubin, Ma, Zhdanov1, Mamun, Hou1, Jiang1, Hou2, Zhdanov2, Bose, Pintu0} and experimentally \cite{Samsonov1, Samsonov2, Melzer, Nosenko1, Nosenko2, Jiang2, Mierk} in dusty plasmas where they can get excited when an object moves with a velocity ($u$) that is greater than the dust acoustic wave velocity ($C_d$). This moving object creates expanding waves (which are circular in 2D) and lead to a V-shaped cone by the superposition of the expanding waves. The geometry of the superposition determines the Mach angle ($\theta$), according to the relation, 
%%%%%%%%%%%%%
\begin{eqnarray}
u/C_d=M = 1/\text{sin}\theta.
\label{vel_cone}
\end{eqnarray}
%%%%%%%%%%%%%%%
Theoretical studies of Mach cones in the layers of dust in Saturn's rings were carried out by Havnes \text{et al.} \cite{Havnes1, Havnes2} who provided estimates of parametric conditions under which they would form. The first experiment on the excitation of Mach cones in a laboratory complex plasma was carried out by Samsonov \textit{et al.} \cite{Samsonov1, Samsonov2}. The Mach cone was excited by a fast moving particle in a stable 2D dust cloud. Unlike the situation in a neutral gas, they observed the formation of two cones - one being a compressive cone arising due to the particles in the stable cloud being pushed by the test particle and the second a rarefactive cone arising due to the particles moving back towards their original positions. Later on a series of molecular-dynamics simulation studies \cite{Jiang1, Hou1,Hou2} were carried out on the excitation of Mach cones that revealed not only the existence of two cones but the possibility of exciting multiple cones because of particle oscillations around their equilibrium lattice position following the passage of the initial disturbance.\par
In this article, we study the dynamics of Mach cones in a dusty plasma by taking into account the correlation effects present in the dust component due to Coulomb interactions between the charged dust particles. For our analysis we employ the generalised hydrodynamic (GH) model that includes this coupling effect between the particles in a phenomenological manner through the incorporation of visco-elasticity and a modified compressibility. This model has been extensively used in the past to investigate the propagation characteristics of linear \cite{Kaw, Mishra} and non-linear waves \cite{Veeresha}, their interaction \cite{Surabhi} and shear-wave Mach cones \cite{Bose} in strongly coupled complex plasmas. Using this model, we obtain analytic expressions for the perturbed dust density and the velocity vector field for the dust fluid from which we construct Mach cone solutions. Our detailed study shows that strong coupling induced effects can have a significant influence on the dynamics and structure of a Mach cone. Specifically we find that the Mach cone angle decreases with an increase in the strong coupling induced compressibility of the medium. We also observe that the arms of a Mach cone structure can break up into several micro-structures at higher coupling strengths between the charged dust particles. Finally the amplitude of the perturbed dust density initially increases to reach its peak value and then monotonically decreases with higher compressibility (corresponding to higher values of the coupling parameter $\Gamma$) of the medium. \par
The paper is organized as follows. In the next section (sec. \ref{sec:theory}), we discuss the governing equations associated with the model used in our investigation. In Section \ref{sec:results}, we discuss our numerical results obtained from the expressions of the  perturbed dust density and mapping of the velocity vector profile in the presence of strong/weak coupling between the particles. Our results are summarized and brief concluding remarks made in section \ref{sec:conclusion}.
\section{Theoretical model}\label{sec:theory}
A variety of approximate models have been used in the past to study \cite{Carini1, Carini2, Postogna, Golden, Berkovsky, Ichimaru} the dynamics of  strongly coupled dusty plasmas. Among these various approaches, one of the most convenient and physically appealing model for investigating strong-coupling effects is the so-called GH model \cite{Boon}. This approach takes into account of the strong correlation effects in the dust dynamics through the introduction of a visco-elastic term in the momentum equation \cite{Kaw} of the dust fluid. The phenomenological GH model has been shown to be valid over a wide range of the coupling parameter, $\Gamma$ ($1\ll \Gamma<\Gamma_c$, where $\Gamma_c$ is the critical value of $\Gamma$ for crystallisation) and has been successfully employed to explain the turnover \cite{Pintu1} in the dispersion relation of a dust acoustic wave and the existence of transverse shear waves \cite{Jyoti, Pintu2} in a strongly coupled dusty plasma medium in liquid state. In the regime where the existence of Mach cone and their dynamics are important, the predominant change due to strong coupling effect is in the dispersion properties. The presence of strong coupling between the particles can modify the compressibility of the medium as seen in the dispersion relation of linear dust acoustic waves. Hence, for studying the Mach cone and their dynamics we retain the effect of compressibility of the fluid along with damping caused by dust neutral collisions but neglect the dissipative contribution due to viscosity. Such an approximation is valid in the so called ``kinetic regime'' where $\omega\tau_m >> 1$, with $\omega$ being the frequency of the mode and $\tau_m$ the relaxation (memory) time due to correlation effects. Thus, under this approximation, the momentum equation for the dust fluid can be written as \cite{Kaw}:  
\begin{eqnarray}
m_d\frac{\partial \mathbf{v_d}(\mathbf{r},t)}{\partial t}&+&m_d\mathbf{v_d}\nabla. \mathbf{v_d}(\mathbf{r},t)=Z_{d}e\nabla \phi(\mathbf{r},t)\nonumber\\
&+&  \mathbf{F}_{EP}-{\frac{\mu T_{d}}{n_{d}}} \frac{\partial {n_d}}{\partial x}.\label{eqn:mom}\\ \nonumber
\end{eqnarray}
The Epstein drag force ($F_{EP}$), arising from the collisions between the background neutral gas molecules and the dust particles, can be expressed as\cite{Epstein},
\begin{eqnarray}
\mathbf{F_{EP}}&=&-\gamma_{EP}m_d\mathbf{v}_d(\mathbf{r},t)\nonumber\\&=&-\delta_{EP}\frac{4\pi}{3}n_nm_nv_na^2\mathbf{v}_d(\mathbf{r},t),
\end{eqnarray}
where, $n_n, v_n, m_n$ are the gas density, thermal velocity and mass of the background neutral gas molecules, respectively. $a$ is the dust particle radius. $\gamma_{EP}$ and $\delta_{EP}$ are the Epstein drag coefficient and a constant which depends on the interaction between the dust particles and gas molecules. \par  
The contribution of compressibility ($\mu$) in the momentum equation (Eq.~(\ref{eqn:mom})) can be expressed in terms of  $\Gamma$ as\cite{Kaw},
\begin{eqnarray}
\mu  = \frac{1}{T_{d}}\left (\frac{\partial P}{\partial n_d}\right)_{T_{d}} = 1+\frac{E(\Gamma)}{3}+\frac{\Gamma} {9}\frac{\partial 
E(\Gamma)}{\partial \Gamma}\label{eqn:mu}
\end{eqnarray}
where $E(\Gamma)$ is the excess internal energy of the system and can be expressed as \cite{Kaw} 
\begin{eqnarray}
\hspace*{-0.5in} E(\Gamma) = -0.89\Gamma + 0.95\Gamma^{1/4} + 0.19\Gamma^{-1/4}-0.81. \label{eqn:gam}
\end{eqnarray}
It is clear from Eq.~\ref{eqn:gam}, in the weakly coupled gaseous phase ($\Gamma < 1 $),  $\mu$ is positive but can become negative as $\Gamma$ increases and one gets into the liquid state. The change in sign of $\mu$ is responsible for the turnover effect in the linear dispersion relation of the dust acoustic wave. \par
The other equations of the model \textit{e.g.}, the continuity equation for the dust component and the Poisson equation can be written are,
\begin{eqnarray}
\frac{\partial n_d(\mathbf{r},t)}{\partial t}&+&\nabla.[n_d(\mathbf{r},t)\mathbf{v_d}(\mathbf{r},t)]=0,\label{eqn:con}
\end{eqnarray}  
\begin{eqnarray}   
\nabla^2\phi(\mathbf{r},t)&=&\frac{e}{\epsilon_0} [Z_dn_d(\mathbf{r},t) + Z_t\delta(\mathbf{r}-\mathbf{u}t)+ n_e(\mathbf{r},t)\nonumber\\ &-& n_i(\mathbf{r},t)].
\label{eqn:pos}
\end{eqnarray}
In the above equations, $n_d(\mathbf{r},t)$ and  $\mathbf{v_d}(\mathbf{r},t)$ represent the instantaneous number density and the velocity vector of the dust fluid, respectively. $m_d$ denotes the mass of dust particles. In our calculations systematic or stochastic dust charge variation are neglected. We also include a $\delta$-function in the Poisson equation (Eq.~\ref{eqn:pos}) to incorporate the presence of a projectile particle which moves with velocity $\mathbf{u}$ and charge number $Z_t$. \par
In the standard fluid description of dusty plasmas for studying low-frequency ($\omega \ll kv_{Te}, kv_{Ti}$) phenomena in the regime where dust dynamics is important, it is customary to treat the electrons and ions as light fluids that can be modeled by Boltzmann distributions and to use the full set of hydrodynamic equations to describe the dynamics of the dust component. The densities of
electrons and ions at temperatures $T_e$ and $T_i$ are, respectively, given by
\begin{eqnarray}
&&n_e = n_{e0} \text{exp}(e\phi(\mathbf{r},t)/k_BT_e),\\
&&n_i = n_{i0} \text{exp}(-e\phi(\mathbf{r},t)/k_BT_i).
 \end{eqnarray}
 The equilibrium electron density $n_{e0}$ and ion density $n_{i0}$ are related to the dust density $n_{d0}$ and the dust charge number $Z_d$ by the charge neutrality condition,
 %$$$$$$$$$$$$$$$$$$$$$$$$$$$$$$$$$$$$$$$
\begin{eqnarray}  
n_{i0}=n_{e0}+n_{d0}Z_d.
\label{eqn:quasi}
\end{eqnarray}
%$$$$$$$$$$$$$$$$$$$$$$$$$$$$$$$$$$$$$$$$
In the perturbed system, assuming a first order approximation, the dynamical variables $n_d(\mathbf{r},t)$, $\mathbf{v_d}(\mathbf{r},t)$, $\phi(\mathbf{r},t)$, $n_e(\mathbf{r},t)$ and $n_i(\mathbf{r},t)$ about the unperturbed states are given by,
\begin{eqnarray}
\nonumber
&n_d(\mathbf{r},t)&=n_{d0}+n_{d1}(\mathbf{r},t),\label{eqn:nd}\\ \nonumber
&\mathbf{v_d}(\mathbf{r},t)&= \mathbf{v_{d1}}(\mathbf{r},t),\label{eqn:vd}\\
&\phi(\mathbf{r},t)&=\phi_1(\mathbf{r},t) ,\label{eqn:phi}\\\nonumber
&n_e(\mathbf{r},t)&=n_{e0}+n_{e0}\left(\frac{e}{k_BT_e}\right)\phi_1(\mathbf{r},t),\label{eqn:ne}\\\nonumber
&n_i(\mathbf{r},t)&=n_{i0}-n_{i0}\left(\frac{e}{k_BT_i}\right)\phi_1(\mathbf{r},t).\label{eqn:ni}
\end{eqnarray}
Using equations (\ref{eqn:quasi}) and (\ref{eqn:phi}) in equations (\ref{eqn:mom}), (\ref{eqn:con}) and (\ref{eqn:pos}) we obtain,
\begin{eqnarray} 
\label{eqn_full} 
\frac{\partial n_{d1}({\bf{r}},t)}{\partial t} &+& n_{d0}\nabla {.{\bf{v}}_{d1}({\bf{r}},t)}=0,\\
\frac{\partial {\bf{v}}_{d1}({\bf{r}},t)}{\partial t}&=&\frac{Z_de}{m_d}\nabla \phi_1({\bf{r}},t)-\gamma_{EP}{\bf{v}}_{d1}({\bf{r}},t)\nonumber\\ &-&\frac{\mu T_d}{m_d n_{d0}} \frac{\partial{n_{d1}}}{\partial {x}},\\
\nabla^2 \phi_1({\bf{r}},t)&=&4\pi e[Z_dn_{d1}({\bf{r}},t)+Z_t\delta({\bf{r}}-{\bf{u}}t)]\nonumber\\&+&\lambda_{D}^{-2}\phi_1({\bf{r}},t).
\end{eqnarray}
We next take a Fourier transform in space and time of the above equations to obtain a set of algebraic equations, where the transform of each quantity is defined as, 
\begin{eqnarray}  
A({\bf{r},t)}=\int{\frac{d{\bf{k}}d\omega}{(2\pi)^4}\tilde{A}({\bf{k}},\omega)e^{i{\bf{k}.r}-i\omega t}}.
\label{eqn_ft}
\end{eqnarray}
The algebraic equations obtained from the Fourier transformed form of Eq~(\ref{eqn_ft}) can be easily solved for
$\tilde{n}_{d1}(\bf{k},\omega),\tilde{{\bf v}}_{d1}(\bf{k},\omega)$ and $\tilde{\phi}_{d1}(\bf{k},\omega)$. Taking the inverse transform
we finally get, 
\begin{eqnarray}  
n_{d1}(\mathbf{r},t)=&\frac{\beta}{(2\pi)^4}\int{d\mathbf{k}d\omega e^{i({\mathbf{k}.r}-\omega t)}}\nonumber\\& \times \frac{[1-\epsilon(k,\omega)]\delta(\omega-\mathbf{k.u})}{\epsilon(k,\omega)}, \label{eqn:density}
\end{eqnarray}  
\begin{eqnarray}  
\mathbf{v}_{d1}(\mathbf{r},t)=&\frac{\beta}{(2\pi)^4n_{d0}}\int{d{\bf{k}}d\omega e^{i(\mathbf{k.r}-\omega t)}}\nonumber\\&\times\frac{\omega}{k^2}  \frac{[1-\epsilon(k,\omega)]\delta(\omega-\mathbf{k.u})}{\epsilon(k,\omega)}{\bf{k}}\label{eqn:velocity},
\end{eqnarray}
where the dielectric function $\epsilon(k,\omega)$ is given by,
\begin{eqnarray}  
\epsilon(k,\omega)=1-\frac{\omega_{pd}^2}{\omega(\omega+i\gamma_{EP})-\frac{\mu T_d k^2}{m_d}}\left(\frac{k^2\lambda_D^2}{k^2\lambda_D^2+1}\right) \label{eqn:die}
\end{eqnarray}
with the dust plasma frequency $\omega_{pd}=(\frac{n_{d0}Z_d^2e^2}{\epsilon_0m_d})^{1/2}$ and the ratio of charges $\beta$= Z$_t$/Z$_d$.\par
Setting $\epsilon(k,\omega)=0$ and $\gamma_0=\gamma_{EP}/\omega_{pd}$, gives the dispersion relation of a linear dust acoustic wave in the presence of strong coupling effects \cite{Kaw},
\begin{equation}
k^2\lambda_D^2=\frac{\omega(\omega+i\gamma_0)-k^2\mu\lambda_D^2}{1-\omega(\omega+i\gamma_0)+k^2\mu\lambda_D^2}.
\label{disp_rel}
\end{equation}
%###############################      Figure 1   ###################################
\begin{figure*}[htb]
\centerline{\hbox{\psfig{file=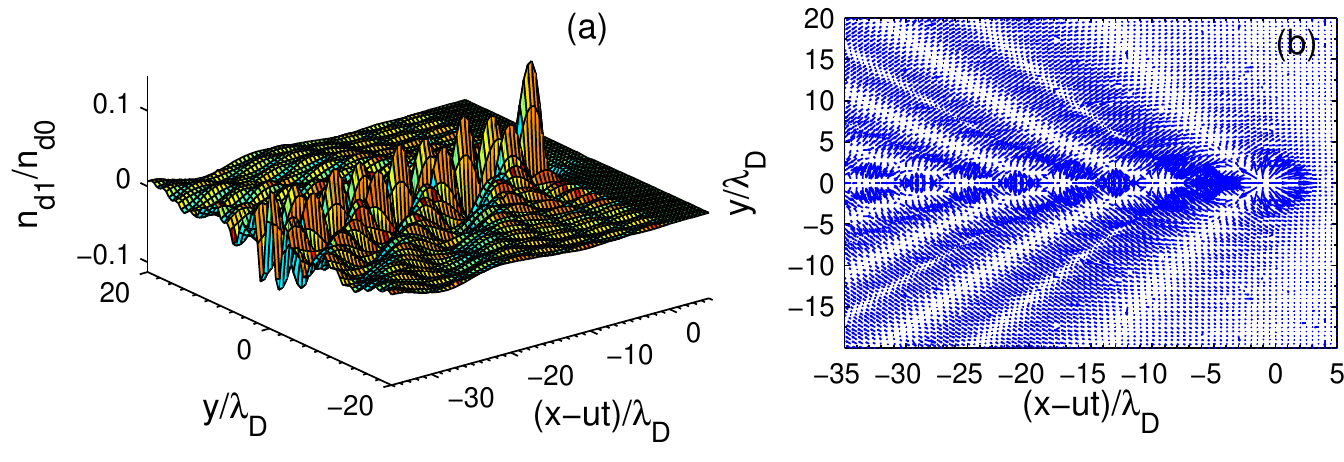,width=1.0\textwidth,angle=0}}}
\caption{(a) Variation of perturbed dust density and (b) velocity vector map of a 2D dust crystal for a given value of Mach number ($M=1.1$), normalised dust neutral collision frequency ($\gamma_0=0.01$) and compressibility ($\mu=-2.0$). }
\label{density_velocity_map}
\end{figure*}
%###################################################################################
Here, we have normalised the frequency of dust acoustic wave by $\omega_{pd}$. 
It is clear from the above equations (Eqs.~(\ref{eqn:density}) and (\ref{eqn:velocity})), that both the normalised perturbed dust density and the velocity are stationary fields in the frame of reference of the projectile particle moving with velocity $\mathbf{u}$.\par
{\color{black}The above dispersion relation (see Eq.~\ref{disp_rel}) implies that a weakly (or strongly) coupled dusty plasma is also a strongly dispersive medium that supports dust acoustic waves of  different phase velocities for different values of the propagation vector ($\mathbf{k}$). As a result, compressional waves excited by the moving particle, will travel with different propagation angles $\theta (\mathbf{k})$ \cite{Dubin, Nosenko2} satisfying the universal Mach-cone angle relation (Eq.~\ref{vel_cone}) and the interaction of these waves can give rise to constructive and destructive interference patterns. Correspondingly, a wake composed of these waves will have a structure containing both a Mach cone and multiple lateral structures. In a dispersionless medium such as  air, such lateral structures are absent and one only observes a single Mach cone being created by an object moving faster than the speed of sound. On the other hand, in a highly dispersive medium like water, lateral wakes are clearly seen behind a ship moving in deep waters, a phenomenon well known as the  \lq Kelvin wedge'. The effect of
dispersion is also clearly seen in our results as will be discussed in the next section.}

\section{Results and Discussions}\label{sec:results}
To study the dynamical characteristics of the Mach cones described by (\ref{eqn:density}) and (\ref{eqn:velocity}) we evaluate these expressions numerically in various parametric regimes and display the results graphically.  The velocity of the particle ($\mathbf{u}$) is expressed in terms of a Mach number, defined as $M=u/C_d$, where $C_d=\lambda_D\omega_{pd}$. 
%############################     Figure 2   #####################################
\begin{figure*}[htb]
\centerline{\hbox{\psfig{file=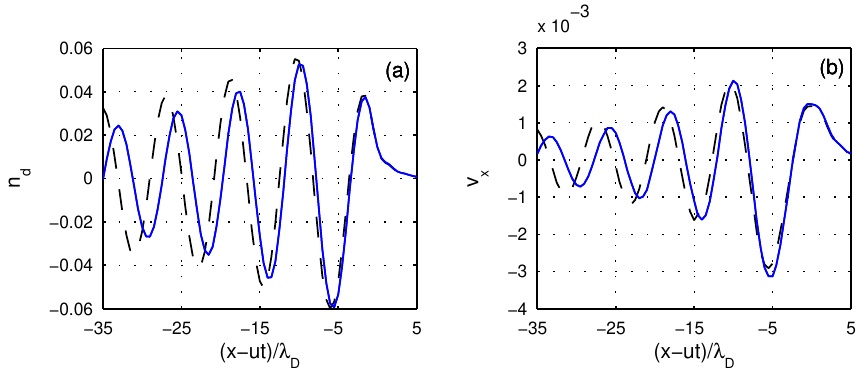,width=1.0\textwidth,angle=0}}}
\caption{Profile of (a)  perturbed dust density along x direction (along which the projectile particle moves) and (b) x-component of velocity for a particular value of $y/\lambda_D (=11.0)$ of fig.~\ref{density_velocity_map}. The peaks on the velocity and density profile indicates the location of maximum velocity and density for $y/\lambda_D=11.0$. The solid and dashed line are for $\mu=0$ and $\mu=-2.0$, respectively.}
\label{vel_den_pro}
\end{figure*}
%###############################################################################
For our numerical evaluation of expressions (\ref{eqn:density}) and (\ref{eqn:velocity}) we have chosen values of the plasma and dust parameters that are close to those of recent experiments \cite{Thomas2}. The plasma density ($n_i$), the electron ($T_e$) and ion ($T_i$) temperatures are chosen as $10^{14}$/m$^3$, $3$ eV and $0.1$ eV, respectively. The dusty plasma parameters \textit{e.g.}, projectile particle charge ($Q_t=Z_te$), average dust charge ($Q_d=Z_de$), dust diameter ($2a$, where a is the radius), inter-particle distance ($d$) are chosen as $10^5e$, $40007e$, $9.0\mu$m and $230\mu$m, respectively. The other parameters, such as dust density ($n_{d0}$), mass of the particles ($m_d$), plasma Debye length ($\lambda_D$) and dust plasma frequency ($\omega_{pd}$), are then estimated using the above plasma and dusty plasma parameters.\par  
%########################## Figure 3 ##################################
\begin{figure*}[ht]
\centerline{\hbox{\psfig{file=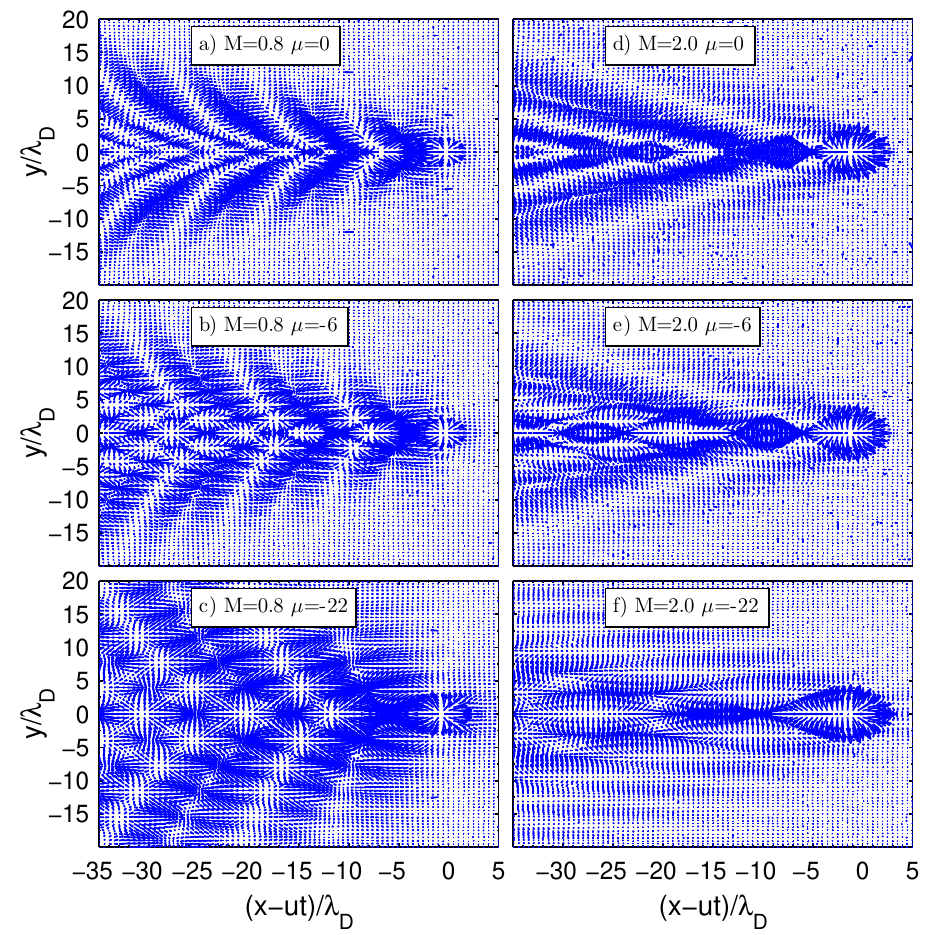,width=1.0\textwidth,angle=0}}}
\caption{Map of the particle velocity vector (${\bf{v}}_{d1}$)  in case of weak damping $\gamma_{0}=0.01$ for (a) $\&$ (d) $\mu=0$, (b) $\&$ (e) $\mu=-6.0$ and (c) $\&$ (f) $\mu=-22.0$. The left panel (a--c) is for $M=0.8$ whereas the right panel (d--f) represents for $M=2.0$.}
\label{qua_1}
\end{figure*}
%###################################################################################
Fig.~\ref{density_velocity_map}  shows the typical normalized perturbed density ($n_{d1}/n_{d0}$) and velocity vector profile map of a lateral wake structure for a given dust-neutral damping coefficient ($\gamma_0=0.01$), Mach number ($M=1.1$) and compressibility ($\mu=-2.0$). The structural properties of the wakes depend on the wave dispersion properties of the medium and the behaviour of wave-particle interaction.  As is evident from both the figures the projectile particle excites V-shaped multiple wake structures which are the results of constructive and destructive interference of waves. From the velocity vector map one notices that all the particles oscillate in the direction perpendicular to the cone wings and parallel to the direction of wave propagation. Hence, it can be concluded that these types of Mach cone structures are composed of longitudinal acoustic waves and, as earlier studies have suggested, the multiple structures arise because of the strong dissipative nature of the dust acoustic waves. \par
%##############################   Figure 4   ###########################
\begin{figure*}[ht]
\centerline{\hbox{\psfig{file=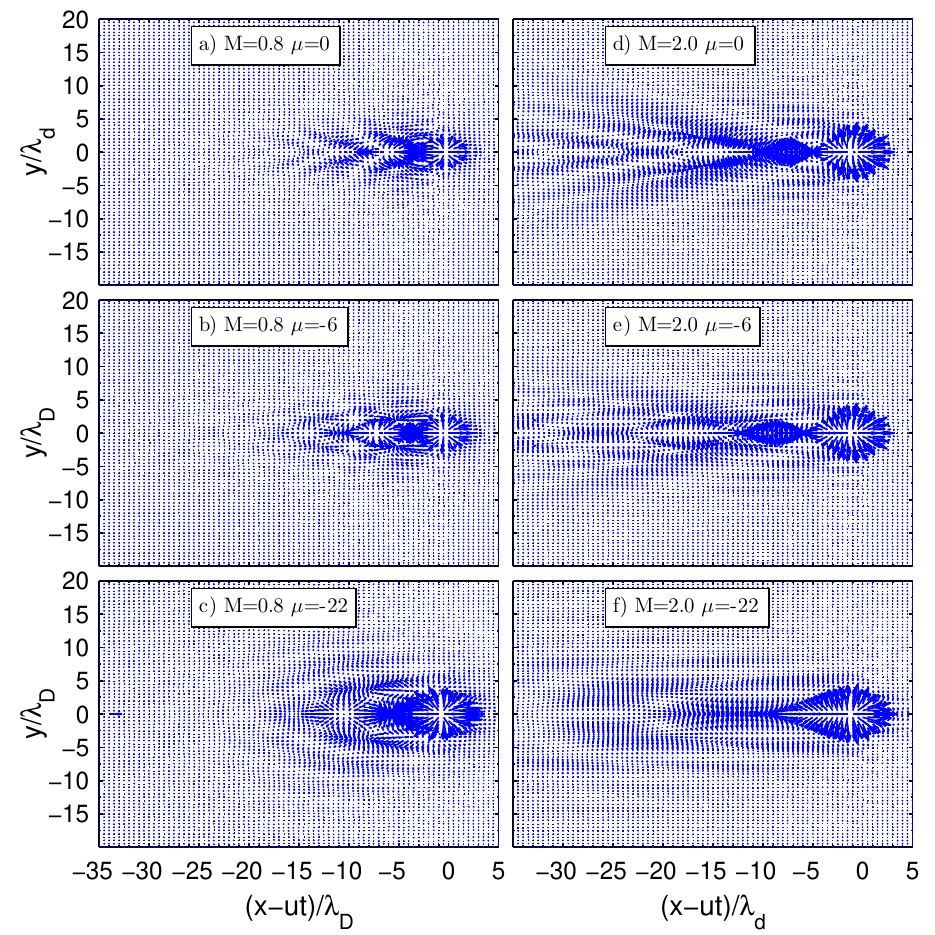,width=1.0\textwidth,angle=0}}}
\caption{Map of the particle velocity vector (${\bf{v}}_{d1}$) in case of strong damping $\gamma_{0}=0.2$ for (a) $\&$ (d) $\mu=0$, (b) $\&$ (e) $\mu=-6.0$ and (c) $\&$ (f) $\mu=-22.0$. The left panel (a--c) is for $M=0.8$ whereas the right panel (d--f) depicts for $M=2.0$.}
\label{qua_2}
\end{figure*}
%###################################################################################
To study the effect of strong coupling on the dynamics of the Mach cone, we next plot the profiles of the amplitude of density and velocity along the X-axis for a particular Y-position in fig.~\ref{vel_den_pro}(a) and \ref{vel_den_pro}(b) respectively for two different values of compressibility $\mu=0$ (solid line) and $\mu=-2.0$ (dashed dotted line). Similar to Fig.~\ref{density_velocity_map}, this figure also shows that the wake structures consist of multiple cones and each cone has a compressive (positive) peak value, with particles pushed in the forward direction, and a rarefactive (negative) peak value, with particles moving back towards their original position. The existence of multiple cones apart from the first one are attributed to the restoring force provided by the electron and ion clouds. It is clear from the figures that the peak positions of these cones appear at different positions for $\mu=0$ and $\mu=-2.0$. It suggests that a higher coupling leads to a higher compressibility of the medium and a lower cone angle. The significant change of cone angle is observed clearly at $x=-35$. \par 
%############################   figure 5 ##########################################
\begin{figure*}[ht]
\centerline{\hbox{\psfig{file=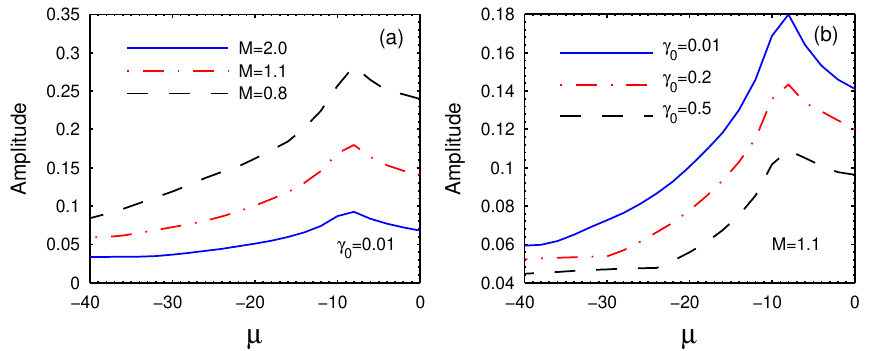,width=1.0\textwidth,angle=0}}}
\caption{Variation of amplitude of normalized perturbed dust density with $\mu$ for different values of (a) $M$ at constant $\gamma_0$ and b) $\gamma_0$ at constant $M$. In figure (a) solid, dash--dotted and dashed lines represent $M$=2.0, 1.1 and 0.8 respectively, at $\gamma_0=0.01$. Whereas in figure (b) solid, dash--dotted and dashed lines represent $\gamma_{0}=0.01$, 0.2 and 0.5, respectively, at $M=1.1$.}
\label{peak_amp}
\end{figure*}
%###################################################################################
Some further snapshots of the Mach cone are shown in fig.~\ref{qua_1} where we have taken the dust-neutral collision frequency ($\gamma_{EP}$) to be very small  compared to the dust oscillation frequency ($\omega_{pd}$) and $\gamma_0=\gamma_{EP}/\omega_{pd}=0.01$. The left panel (a)--(c) of fig.~\ref{qua_1} depicts the velocity vector profile maps for a given Mach number (corresponding to the same projectile particle velocity), $M=0.8$ but for different values of $\mu$: (a) $\mu=0$, (b) $\mu=-6.0$ and (c) $\mu=-22.0$. The corresponding values of $\Gamma$ for these $\mu$ values can be found out from the Eq.~\ref{eqn:gam}. The right panel (d)--(f) of this figure corresponds to different values of Mach number \textit{i.e.} $M=2.0$ but for same set of compressibility. It is clear from the left panel of the figure that the structural changes are significant as one increases the rigidity of the medium by changing the coupling factor for a constant value of $M$ and $\gamma_0$. As discussed above in the context of fig.~\ref{vel_den_pro}, the cone angle decreases with the increase of the compressibility of the medium. Increase of Mach number, results in the decrease of the opening angle of the Mach cones which satisfy the Mach-cone-angle relationship.  Additionally, the cone structures become turbulent when $\mu$ reaches towards its higher values and the wings of the cone breaks into small micro-structures.  The wake structures even get washed out at higher values of $\mu$ as shown in fig.~\ref{qua_1}.  It is also to be noted that the Mach cone angle decreases when Mach number increases from $M=0.8$ to $M=2.0$ (see right panel of fig.~\ref{qua_1}). This can also be explained by the Mach-cone-angle relationship. It is also observed that the number of multiple structures reduces with the increase of Mach number for a given value of $\mu$ and $\gamma_0$.  With the decrease of the Mach number the structures are also found to be more unstable for the same value of $\mu$. \par
In order to study the role of damping on the dynamics of lateral wake structures,  we next increase $\gamma_0$ to 0.2 and plot the velocity vector maps in fig.~\ref{qua_2} for two different values of Mach numbers, $M=0.8$ (left panel) and $M=2.0$ (right panel) for three different values of $\mu$ similar to fig.~\ref{qua_1}. It is observed that the wakes are now strongly damped because of the frequent collisions of the dust particles with the background neutrals. The oscillations in the wake region are smoothed out due to the decrease of the damping length, $l_d=C_d/\gamma_{EP}$ \cite{Samsonov2, Nosenko1, Hou1, Pintu0}. The figure also suggests that the high collisionality reduces the multiple wake structures down to two or even one irrespective of Mach number and the compressibility of the medium. But size of the locus from where the wakes emanate increases with the increase of $\mu$ value. For higher values of $\mu$, it is seen that the circular locus becomes elliptical.\par
{\color{black} We next discuss the influence of wave dispersion on the nature of the wake patterns observed in our model calculations. As pointed out in \cite{Dubin, Nosenko2}, wake structures can assume complex forms due to the presence of dispersion in the medium. The effect is more pronounced when $M$ is close to unity whereas for $M >> 1$ or $ M << 1$ the structures are much simpler often consisting of a single Mach cone or a wake structure. We see clear evidence of this in our present investigations as well. The left panels of fig.~\ref{qua_1}  and fig.~\ref{qua_2} are plotted for $M=0.8$ (barely subsonic for compressional waves) which show the wakes to consist of multiple complicated lateral structures. Similar behavior  can also be seen in the case of fig.~\ref{density_velocity_map} which is a plot for $M=1.1$. On the other hand, the wake structures shown in the velocity map of the right panels of fig.~\ref{qua_1}  and fig.~\ref{qua_2}, that are plotted for $M=2.0$ (in strongly supersonic case) contain fewer number of wakes (maximum two, often reducing to one). It is also to be noted that for $M <1$ the Mach cone should disappear leaving only wake patterns. This is also seen in the velocity maps for $M=0.8$ (see left panels of fig.~\ref{qua_1}  and fig.~\ref{qua_2}). } \par
Fig.~\ref{peak_amp} illustrates the variation of maximum amplitudes of perturbed dust density of a 2D plasma crystal with the compressibility of the medium for a set of Mach numbers and dust-neutral collision rate. Fig.~\ref{peak_amp}(a) represents the same when the dust neutral collision frequency remains unchanged at $\gamma_0=0.01$ and the Mach number changes from $0.8$ to $2.0$. It is seen that for a given Mach number, the amplitude increases with the increase of $\mu$ and reaches its maximum value and then it decreases monotonically towards zero value. The position of the maxima is found to be around $\mu \approx -8.0$. For a given value of $\mu$, it is also to be noted that the amplitude increases with the decrease of Mach number. Variation of maximum amplitude as a  function of $\mu$ shows a similar trend when we change the dust-neutral collision frequency from 0.01 to 0.5 at a constant value of Mach number (M=1.1) as shown in fig.~\ref{peak_amp}(b). The peak amplitude of perturbed dust density is higher for lower $\gamma_0$ for a given value of Mach number and the medium of same rigidity.  
\section{Conclusion}
\label{sec:conclusion}
We have investigated the dynamics of Mach cones  in a strongly coupled unmagnetised 2D complex plasma. We have employed the Generalized-Hydrodynamic equations to model the dust dynamics and incorporated the strong coupling induced dispersive effects through the modification in the compressibility of the medium. To excite the Mach cone we have taken into account the effect of a projectile particle  in the Poisson equation and solved the fluid equations to obtain expressions for the perturbed dust density and the velocity vector profile. Numerical plots of these expressions for relevant plasma/dust parameters show the existence of compressional oscillatory wake structures. We have studied the changes in the characteristics of such Mach cones as a function of the dust neutral collision frequency as well as strong coupling induced changes in the compressibility of the medium. We find that with the increase of compressibility the opening angle of the Mach cone decreases and additionally the wake structure gets turbulent through a breakup of its wings into micro structures. At very high value of $\mu$ the wing structures get washed out. A quantitative study shows that the peak amplitude of the perturbed dust density first increases with the compressibility of the medium to reach a peak value and then it monotonically decreases. For a given value of $\mu$ the amplitude increases with the decrease of Mach number and the dust neutral collision rate. These features should be observable in laboratory experiments and can provide a direct measure of strong coupling induced modifications of the system compressibility. Although Mach cones have been studied for a long time but their evolution in the presence of strong coupling effects have not received adequate attention and our present theoretical results can help provide directions for future experiments as well as simulation studies.  \par
\begin{acknowledgements}
One of the authors, Sangeeta, acknowledges the support from the Institute for Plasma Research (IPR) for carrying out this research work during the Summer School Program-2014 at IPR.
\end{acknowledgements}
%+++++++++++++++++++++++++++++++++++++++++
%\section*{References}

\end{document}